\documentclass[10pt,a4paper]{article}

\usepackage[utf8]{inputenc}
\usepackage[english]{babel}

\begin{document}
\large

\newpage
\begin{center}
\LARGE{\bf Vector Currents of Massive Neutrinos of 
\\an Electroweak Nature}
\end{center}
\vspace{0.1cm}
\begin{center}
{\bf Rasulkhozha S. Sharafiddinov}
\end{center}
\vspace{0.1cm}
\begin{center}
{\bf Institute of Nuclear Physics, Uzbekistan Academy of Sciences,
\\Tashkent, 100214 Ulugbek, Uzbekistan}
\end{center}
\vspace{0.1cm}

\begin{center}
{\bf Abstract}
\end{center}

The mass of an electroweakly interacting neutrino consists of the electric
and weak parts responsible for the existence of its charge, charge radius,
and magnetic moment. Such connections explain the formation of paraneutrinos,
for example, at the polarized neutrino electroweak scattering by spinless
nuclei. We derive the structural equations that relate the self-components
of mass to charge, charge radius, and magnetic moment of each neutrino as a
consequence of unification of fermions of a definite flavor. They indicate
the availability of neutrino universality and require following its logic
in a constancy law dependence of the size implied from the multiplication of
a weak mass of neutrino by its electric mass. According to this principle, all
Dirac neutrinos of a vector nature, regardless of the difference in their masses,
have the same charge, an identical charge radius, as well as an equal magnetic 
moment. Thereby, the possibility appears of establishing the laboratory 
limits of weak masses of the investigated types of neutrinos. Finding estimates 
show clearly that the earlier measured properties of these particles may 
testify in favor of the unified mass structure of their interaction with 
any of the corresponding types of gauge fields.

\vspace{0.8cm}
\noindent
{\bf 1. Introduction}
\vspace{0.4cm}

A notion about neutrino oscillation introduced by Pontecorvo [1] may be connected with 
a principle, according to which, any neutrinos of Majorana types [2] must have their own 
Dirac neutrino of true neutrality. Such a nonclassical correspondence, regardless of 
the nature of the C-invariant Dirac neutrino, expresses the idea of a coexistence 
law [3] of C-noninvariant neutrinos of Dirac and Majorana types. 

From this point of view, each of earlier experiments [4,5] about mixing angles may serve as 
the source of facts confirming the existence in all truly neutral particles of a kind of C-odd electric charge [6,7] responsible for the flavor symmetrical mode of neutrino oscillations. The Coulomb transitions of these types can explain the absence of vector currents of truly neutral neutrinos and the availability of an axial-vector nature of their mass [8,9]. 

In classical electrodynamics, it has usually been assumed that all inertial mass of a particle 
is equal to its electric mass [10,11]. This implies that elementary objects with Coulomb behavior have neither weak, strong, nor any other type of interaction. This is, however, valid only for 
those particles in which mass is absent.

One such an object may, according to earlier presentations, be a Dirac neutrino of a 
C-invariant nature. But unlike the first-initial two component theory [12] of the neutrino,
its logically consistent development [13] gives the possibility to relate the mass, $m_{\nu_{l}},$ 
to charge, $F_{1\nu_{l}}(q^{2}),$ and magnetic, $F_{2\nu_{l}}(q^{2}),$ form factors of this particle, 
$(l=e,\mu,\tau,...),$ as a consequence of the equality of the interaction cross sections [6,7] 
with the field of emission of both types of vector, $V_{\nu_{l}},$ currents. For the case when 
their independent, $f_{i\nu_{l}}(0),$ parts respond to the process, the latter is reduced to 
the following prediction of flavor symmetry:
\begin{equation}
f_{1\nu_{l}}(0)-2m_{\nu_{l}}f_{2\nu_{l}}(0)=0
\label{1}
\end{equation}
At the same time, it is clear that functions $F_{i\nu_{l}}(q^{2})$ depending on the momentum
transfer square $q^{2}$ include not only static but also dynamic components
\begin{equation}
F_{i\nu_{l}}(q^{2})=f_{i\nu_{l}}(0)+R_{i\nu_{l}}(q^{2})+...
\label{2}
\end{equation}
where $f_{i\nu_{l}}(0),$ as will be seen later, give the dimensional sizes of the neutrino electric charge and magnetic dipole moment. The second terms $R_{i\nu_{l}}(q^{2})$ characterize a connection of form factors with a particle electromagnetic radius.

Insofar as the function $R_{1\nu_{l}}(q^{2})$ is concerned, it describes the interaction 
between the charge, $r_{\nu_{l}},$ radius of the neutrino and the field of emission
of the photon: $R_{1\nu_{l}}(q^{2})=(q^{2}/6)<r^{2}_{\nu_{l}}>.$

It is interesting, however, that any dipole arises as a result of a kind of charge [14]. 
Therefore, if it turns out that each neutrino having a C-even or a C-odd charge possesses 
a mass of a vector or an axial-vector nature [8,9], from the point of view of any 
of them, it should be expected that the term $f_{2\nu_{l}}(0)$ is the dipole of a vector 
C-invariant charge that does not coincide with the dipole that arises as a consequence of 
an axial-vector C-noninvariant charge. 

Such a unified principle corresponds, in the limit of the Dirac neutrino of a vector nature, to
the previously mentioned connection of C-invariant, $f_{i\nu_{l}}(0),$ currents and, consequently,
there exists a certain latent dependence between the mass of the neutrino and its charge.

This would seem to contradict charge quantization. As was, however, noted for the first time by the author [15], to any type of electrically charged particle corresponds a kind of magnetically charged monoparticle. In a given situation, each mononeutrino responds to quantization of the electric
charges of all neutrinos and vice versa.

One can also use an arbitrary charge as an example, introduction [16] of which into the framework 
of the standard electroweak theory [17-19] is not excluded.

At the availability of the suggested connection, the conservation of charge
in the decays of the neutron, muon, tau lepton and in other reactions with 
neutrinos must lead to a formation in the field of emission of dileptons
of a definite flavor [20].

Another characteristic moment is the mass structure [3] of gauge invariance. It states that 
the existence of charge and its radius in a massive neutrino is incompatible with the absence 
of gauge symmetry.

In the presence of a purely electric part of mass, the expected structure of
$f_{1\nu_{l}}(0)$ encounters the condition of the steadiness of charge distribution
in a neutrino and requires explanation from the point of view of the interratio
of the most diverse types of intraneutrino forces. For this we must at first
recall the mass-charge duality [21], according to which, each of the
Coulomb, weak, and unelectroweak charges says about the existence in nature
of a kind of inertial mass. Therefore, a neutrino with electroweak behavior
can have not only electric [10,11] but also weak [22] masses.

Thus, all the mass, $m_{\nu_{l}},$ and charge, $e_{\nu_{l}},$ of the neutrino
coincide with its electroweakly united $(EW)$ mass and charge
\begin{equation}
m_{\nu_{l}}=m_{\nu_{l}}^{EW}=m_{\nu_{l}}^{E}+m_{\nu_{l}}^{W}
\label{3}
\end{equation}
\begin{equation}
e_{\nu_{l}}=e_{\nu_{l}}^{EW}=e_{\nu_{l}}^{E}+e_{\nu_{l}}^{W}
\label{4}
\end{equation}
possessing the Coulomb $(E)$ and weak $(W)$ components. They constitute
the intraneutrino harmony of the four types of forces [23].

For the further substantiation of the legality of such a procedure one must build the functions $f_{i\nu_{l}}(0)$ and $<r^{2}_{\nu_{l}}>$ in the neutrino mass structure dependence. From this purpose, we investigate here the behavior of elastic scattering of longitudinal polarized 
neutrinos of a C-invariant nature on a spinless nucleus as a consequence of the availability 
of the electric, $m_{\nu_{l}}^{E},$ and weak, $m_{\nu_{l}}^{W},$ masses, and also of charge,
charge radius, and magnetic moment of incoming fermions of vector weak, $V_{\nu_{l}},$ currents.

\vspace{0.8cm}
\noindent
{\bf 2. Unity of neutrino vector electroweak interaction structural parts}
\vspace{0.4cm}

The matrix elements of the transitions noted earlier [24] in the limit of one-boson exchange include the following current parts:
$$M^{E}_{fi}=\frac{4\pi\alpha}{q_{E}^{2}}\overline{u}(p_{E}',s')
\{\gamma_{\mu}[f_{1\nu_{l}}^{E}(0)+
\frac{1}{6}q_{E}^{2}<r^{2}_{\nu_{l}}>_{E}]$$
\begin{equation}
-i\sigma_{\mu\lambda}q_{\lambda E}f_{2\nu_{l}}^{E}(0)\}
u(p_{E},s)<f|J_{\mu}^{\gamma}(q_{E})|i>
\label{5}
\end{equation}
\begin{equation}
M^{W}_{fi}=
\frac{G_{F}}{\sqrt{2}}\overline{u}(p_{W}',s')\gamma_{\mu}g_{V_{\nu_{l}}}^{*}
u(p_{W},s)<f|J_{\mu}^{Z}(q_{W})|i>
\label{6}
\end{equation}
where $\nu_{l}=\nu_{lL,R}({\bar \nu_{lR,L}});$ $q_{E}=p_{E}-p_{E}';$
$q_{W}=p_{W}-p_{W}';$ $p_{E}(p_{W})$ and $p_{E}'(p_{W}')$ correspond in the
Coulomb (weak) scattering to the four-momenta of initial and final neutrinos
of the definite helicities $s$ and $s';$ $J_{\mu}^{\gamma}$ and
$J_{\mu}^{Z}$ describe the target nucleus currents at the emission
of virtual photons and $Z$-bosons; $g_{V_{\nu_{l}}}^{*}$ distinguishes
from $g_{V_{\nu_{l}}},$ namely, from the neutrino weak interaction vector
component constant by a multiplier $(1/\sin\theta_{W}),$ which arises
only in the case where $e_{l}^{E}=1$ when
\begin{equation}
e_{\nu_{l}}^{E}=e_{\nu_{l}}^{W}\sin\theta_{W}
\label{7}
\end{equation}
The index $E$ in $f_{i\nu_{l}}^{E}$ and $<r^{2}_{\nu_{l}}>_{E}$ implies the
availability of a connection between these characteristics of the neutrino
and an electric, $m_{\nu_{l}}^{E},$ part of its all rest mass. We see in
addition that in the case of exchange by the $Z$-boson, only the weak,
$m_{\nu_{l}}^{W},$ component of mass is responsible for the scattering.

A neutrino itself possesses simultaneously both electric [10,11] and weak [22] masses. This 
in turn leads to those processes that originate at the expense of the mixedly interference 
$(I)$ interaction [24]
$$ReM^{E}_{fi}M^{*W}_{fi}=
\frac{4\pi\alpha G_{F}}{\sqrt{2}q_{I}^{2}}
Re\Lambda_{I}\Lambda_{I}'\{\gamma_{\mu}
[f_{1\nu_{l}}^{I}(0)+
\frac{1}{6}q_{I}^{2}<r^{2}_{\nu_{l}}>_{I}]$$
\begin{equation}
-i\sigma_{\mu\lambda}q_{\lambda I}f_{2\nu_{l}}^{I}(0)\}
\gamma_{\mu}g_{V_{\nu_{l}}}^{*}
J_{\mu}^{\gamma}(q_{I})J_{\mu}^{Z}(q_{I})	
\label{8}
\end{equation}
where the currents $f_{i\nu_{l}}^{I}$ and $<r^{2}_{\nu_{l}}>_{I}$ appear in the mass and 
charge structure dependence. Here one must keep also in mind that
$$q_{I}=p_{I}-p_{I}'$$
$$\Lambda_{I}=u(p_{I},s)\overline{u}(p_{I},s)$$
$$\Lambda_{I}'=u(p_{I}',s')\overline{u}(p_{I}',s')$$
Among them $p_{I}$ and $p_{I}'$ express the four-momenta of the neutrino
before and after the interaction with an interference field of emission
of the photon and weak boson. Thereby, they describe a situation when an
interference mass of the neutrino, $m_{\nu_{l}}^{I},$ does not coincide with
its all rest mass, $m_{\nu_{l}}^{EW}.$ Such a distinction between the sizes
of $m_{\nu_{l}}^{I}$ and $m_{\nu_{l}}^{EW}$ appears in the difference of
electric, $m_{\nu_{l}}^{E},$ and weak, $m_{\nu_{l}}^{W},$ parts of the neutrino
mass. This connection, similarly to ratio (\ref{7}), will correspond in
nature to the electroweak unification at a more fundamental dynamical level.

For spinless nuclei of the electric, $Z,$ and weak, $Z_{W},$ charges and of all C-even types 
of longitudinal polarized neutrinos, the investigated process cross section, according 
to (\ref{5})-(\ref{8}) and the standard definition
\begin{equation}
\frac{d\sigma_{EW}(s,s')}{d\Omega}=
\frac{1}{16\pi^{2}}|M^{E}_{fi}+M^{W}_{fi}|^{2}
\label{9}
\end{equation}
may be written as
\begin{equation}
d\sigma_{EW}^{V_{\nu_{l}}}(\theta_{EW},s,s')=
d\sigma_{E}^{V_{\nu_{l}}}(\theta_{E},s,s')+
d\sigma_{I}^{V_{\nu_{l}}}(\theta_{I},s,s')+
d\sigma_{W}^{V_{\nu_{l}}}(\theta_{W},s,s')
\label{10}
\end{equation}
where $\theta_{EW}$ is the neutrino scattering angle in an electroweakly
united $(EW)$ interaction.

The purely Coulomb contributions are equal to
$$\frac{d\sigma_{E}^{V_{\nu_{l}}}(\theta_{E},s,s')}{d\Omega}=
\frac{1}{2}\sigma^{E}_{o}(1-\eta^{2}_{E})^{-1}\{(1+ss')[f_{1\nu_{l}}^{E}$$
$$-\frac{2}{3}<r^{2}_{\nu_{l}}>_{E}(m_{\nu_{l}}^{E})^{2}\gamma_{E}^{-1}]^{2}$$
$$+\eta^{2}_{E}(1-ss')[(f_{1\nu_{l}}^{E}-
\frac{2}{3}<r^{2}_{\nu_{l}}>_{E}(m_{\nu_{l}}^{E})^{2}\gamma_{E}^{-1})^{2}$$
\begin{equation}
+4(m_{\nu_{l}}^{E})^{2}(1-\eta^{-2}_{E})^{2}(f_{2\nu_{l}}^{E})^{2}]
tg^{2}\frac{\theta_{E}}{2}\}
F_{E}^{2}(q_{E}^{2})
\label{11}
\end{equation}
To the interference scattering responds the expression
$$\frac{d\sigma_{I}^{V_{\nu_{l}}}(\theta_{I},s,s')}{d\Omega}=
\frac{1}{2}\rho_{I}\sigma^{I}_{o}(1-\eta^{2}_{I})^{-1}
g_{V_{\nu_{l}}}\{(1+ss')[f_{1\nu_{l}}^{I}$$
$$-\frac{2}{3}<r^{2}_{\nu_{l}}>_{I}(m_{\nu_{l}}^{I})^{2}\gamma_{I}^{-1}]+
\eta^{2}_{I}(1-ss')[f_{1\nu_{l}}^{I}$$
\begin{equation}
-\frac{2}{3}<r^{2}_{\nu_{l}}>_{I}(m_{\nu_{l}}^{I})^{2}\gamma_{I}^{-1}]
tg^{2}\frac{\theta_{I}}{2}\}F_{I}(q_{I}^{2})
\label{12}
\end{equation}
The cross section explained by the weak interaction (\ref{6}) has the form
$$\frac{d\sigma_{W}^{V_{\nu_{l}}}(\theta_{W},s,s')}{d\Omega}=
\frac{G_{F}^{2}(m_{\nu_{l}}^{W})^{2}}{16\pi^{2}\sin^{2}\theta_{W}}
g_{V_{\nu_{l}}}^{2}\{\eta_{W}^{-2}(1+ss')\cos^{2}\frac{\theta_{W}}{2}$$
\begin{equation}
+(1-ss')\sin^{2}\frac{\theta_{W}}{2}\}F_{W}^{2}(q_{W}^{2})
\label{13}
\end{equation}
Here we have used the relations:
$$\sigma_{o}^{E}=\frac{\alpha^{2}}{4(m_{\nu_{l}}^{E})^{2}}
\frac{\gamma_{E}^{2}}{\alpha_{E}} \, \, \, \,
\rho_{I}=-\frac{2G_{F}(m_{\nu_{l}}^{I})^{2}}
{\pi\sqrt{2}\alpha\sin\theta_{W}}\gamma_{I}^{-1}$$
$$\sigma_{o}^{I}=\frac{\alpha^{2}}{4(m_{\nu_{l}}^{I})^{2}}
\frac{\gamma_{I}^{2}}{\alpha_{I}} \, \, \, \,
\alpha_{K}=\frac{\eta^{2}_{K}}{(1-\eta^{2}_{K})
\cos^{2}(\theta_{K}/2)}$$
$$\gamma_{K}=\frac{\eta^{2}_{K}}{(1-\eta^{2}_{K})
\sin^{2}(\theta_{K}/2)} \, \, \, \,
\eta_{K}=\frac{m_{\nu_{l}}^{K}}{E_{\nu_{l}}^{K}}$$
$$F_{E}(q_{E}^{2})=ZF_{c}(q_{E}^{2}) \, \, \, \,
F_{I}(q_{I}^{2})=ZZ_{W}F_{c}^{2}(q_{I}^{2})$$
$$F_{W}(q_{W}^{2})=Z_{W}F_{c}(q_{W}^{2}) \, \, \, \,
q_{K}^{2}=-4(m_{\nu_{l}}^{K})^{2}\gamma_{K}^{-1}$$
$$Z_{W}=\frac{1}{2}\{\beta_{V}^{(0)}(Z+N)+\beta_{V}^{(1)}(Z-N)\}$$
$$A=Z+N \, \, \, \, M_{T}=\frac{1}{2}(Z-N)$$
$$\beta_{V}^{(0)}=-2\sin^{2}\theta_{W}
\, \, \, \, \beta_{V}^{(1)}=\frac{1}{2}-2\sin^{2}\theta_{W}$$
$$g_{V_{\nu_{l}}}=-\frac{1}{2}+2\sin^{2}\theta_{W} \, \, \, \, K=E,I,W$$
where $\theta_{K}$ denote the neutrino Coulomb, interference, and weak
scattering angles at the energies $E_{\nu_{l}}^{K},$ the functions
$F_{c}(q_{K}^{2})$ are responsible in these processes for charge $(F_{c}(0)=1)$
distribution of a nucleus with an isospin $T$ and its projection $M_{T};$
$\beta_{V}^{(0)}$ and $\beta_{V}^{(1)}$ characterize the constants of
hadronic vector weak current isoscalar and isovector parts.

The presence of self-interference terms $(f_{i\nu_{l}}^{E})^{2}$ and
$<r_{\nu_{l}}^{4}>_{E}$ in (\ref{11}) is explained by the formation of
the left- or right-handed [20] paraneutrinos
\begin{equation}
(\nu_{lL}, {\bar \nu_{lR}}) \, \, \, \,
(\nu_{lR}, {\bar \nu_{lL}})	
\label{14}
\end{equation}
Their appearance in the nuclear Coulomb field can also be explained by the
contribution $f_{1\nu_{l}}^{E}<r_{\nu_{l}}^{2}>_{E}$ of the mixed interference
between the interactions with a photon of the neutrino charge and charge radius.
They of course appear also at the expense of weak currents.
In the latter case from (\ref{13}) and its structural components,
$g_{V_{\nu_{l}}}^{2},$ we are led to a correspondence principle
that the nature of difermions depends on an interaction type.
Therefore, the availability of mixedly interference contributions
$g_{V_{\nu_{l}}}f_{1\nu_{l}}^{I}$ and $g_{V_{\nu_{l}}}<r_{\nu_{l}}^{2}>_{I}$
in the scattering cross section (\ref{12}) can also confirm that
the existence of paraneutrinos with electroweak behavior is, by itself,
not excluded.

Here it is relevant to note that (\ref{9}) redoubles the value of mixedly
interference terms. But the number of difermions and those phenomena that
lead to their formation coincide. Such a symmetry explains the separation of 
any type of the mixedly interference cross section into the two equal parts.

Thus, if we sum each of (\ref{11})-(\ref{13}) over $s',$ one can write
(\ref{10}) in the form
\begin{equation}
d\sigma_{EW}^{V_{\nu_{l}}}(\theta_{EW},s)=
d\sigma_{E}^{V_{\nu_{l}}}(\theta_{E},s)+
\frac{1}{2}d\sigma_{I}^{V_{\nu_{l}}}(\theta_{I},s)+
\frac{1}{2}d\sigma_{I}^{V_{\nu_{l}}}(\theta_{I},s)+
d\sigma_{W}^{V_{\nu_{l}}}(\theta_{W},s)
\label{15}
\end{equation}
where the purely Coulomb scattering cross section behaves as
$$d\sigma_{E}^{V_{\nu_{l}}}(\theta_{E},s)=
d\sigma_{E}^{V_{\nu_{l}}}(\theta_{E},f_{1\nu_{l}}^{E},s)+
\frac{1}{2}d\sigma_{E}^{V_{\nu_{l}}}
(\theta_{E},f_{1\nu_{l}}^{E},<r^{2}_{\nu_{l}}>_{E},s)$$
$$+\frac{1}{2}d\sigma_{E}^{V_{\nu_{l}}}
(\theta_{E},f_{1\nu_{l}}^{E},<r^{2}_{\nu_{l}}>_{E},s)$$
\begin{equation}
+d\sigma_{E}^{V_{\nu_{l}}}(\theta_{E},<r^{2}_{\nu_{l}}>_{E},s)+
d\sigma_{E}^{V_{\nu_{l}}}(\theta_{E},f_{2\nu_{l}}^{E},s)
\label{16}
\end{equation}

$$\frac{d\sigma_{E}^{V_{\nu_{l}}}(\theta_{E},f_{1\nu_{l}}^{E},s)}{d\Omega}=
\frac{d\sigma_{E}^{V_{\nu_{l}}}(\theta_{E},f_{1\nu_{l}}^{E},s'=s)}{d\Omega}+
\frac{d\sigma_{E}^{V_{\nu_{l}}}(\theta_{E},f_{1\nu_{l}}^{E},s'=-s)}{d\Omega}$$
\begin{equation}
=\sigma^{E}_{o}(1-\eta^{2}_{E})^{-1}(1+\eta_{E}^{2}tg^{2}
\frac{\theta_{E}}{2})(f_{1\nu_{l}}^{E})^{2}F_{E}^{2}(q^{2}_{E})
\label{17}
\end{equation}

$$\frac{d\sigma_{E}^{V_{\nu_{l}}}
(\theta_{E},f_{1\nu_{l}}^{E},<r^{2}_{\nu_{l}}>_{E},s)}{d\Omega}=
\frac{d\sigma_{E}^{V_{\nu_{l}}}
(\theta_{E},f_{1\nu_{l}}^{E},<r^{2}_{\nu_{l}}>_{E},s'=s)}
{d\Omega}$$
$$+\frac{d\sigma_{E}^{V_{\nu_{l}}}
(\theta_{E},f_{1\nu_{l}}^{E},<r^{2}_{\nu_{l}}>_{E},s'=-s)}
{d\Omega}$$
$$=-\frac{2}{3}(m_{\nu_{l}}^{E})^{2}\gamma_{E}^{-1}\sigma^{E}_{o}
(1-\eta^{2}_{E})^{-1}$$
\begin{equation}
\times (1+\eta_{E}^{2}tg^{2}\frac{\theta_{E}}{2})
f_{1\nu_{l}}^{E}<r^{2}_{\nu_{l}}>_{E}F_{E}^{2}(q^{2}_{E})
\label{18}
\end{equation}

$$\frac{d\sigma_{E}^{V_{\nu_{l}}}
(\theta_{E},<r^{2}_{\nu_{l}}>_{E},s)}{d\Omega}=
\frac{d\sigma_{E}^{V_{\nu_{l}}}
(\theta_{E},<r^{2}_{\nu_{l}}>_{E},s'=s)}{d\Omega}$$
$$+\frac{d\sigma_{E}^{V_{\nu_{l}}}
(\theta_{E},<r^{2}_{\nu_{l}}>_{E},s'=-s)}{d\Omega}$$
$$=\frac{4}{9}(m_{\nu_{l}}^{E})^{4}\gamma_{E}^{-2}\sigma^{E}_{o}
(1-\eta^{2}_{E})^{-1}$$
\begin{equation}
\times (1+\eta_{E}^{2}tg^{2}\frac{\theta_{E}}{2})
<r^{4}_{\nu_{l}}>_{E}F_{E}^{2}(q^{2}_{E})
\label{19}
\end{equation}

$$\frac{d\sigma_{E}^{V_{\nu_{l}}}(\theta_{E},f_{2\nu_{l}}^{E},s)}{d\Omega}=
\frac{d\sigma_{E}^{V_{\nu_{l}}}(\theta_{E},f_{2l}^{E},s'=-s)}{d\Omega}$$
\begin{equation}
=4(m_{\nu_{l}}^{E})^{2}\eta^{-2}_{E}\sigma^{E}_{o}(1-\eta^{2}_{E})^{2}
(f_{2\nu_{l}}^{E})^{2}F_{E}^{2}(q^{2}_{E})tg^{2}\frac{\theta_{E}}{2}
\label{20}
\end{equation}
The second term in (\ref{15}) corresponds to the electroweak interference
process and becomes equal to
\begin{equation}
d\sigma_{I}^{V_{\nu_{l}}}(\theta_{I},s)=
d\sigma_{I}^{V_{\nu_{l}}}(\theta_{I},g_{V_{\nu_{l}}},f_{1\nu_{l}}^{I},s)+
d\sigma_{I}^{V_{\nu_{l}}}
(\theta_{I},g_{V_{\nu_{l}}},<r^{2}_{\nu_{l}}>_{I},s)
\label{21}
\end{equation}

$$\frac{d\sigma_{I}^{V_{\nu_{l}}}
(\theta_{I},g_{V_{\nu_{l}}},f_{1l}^{I},s)}{d\Omega}=
\frac{d\sigma_{I}^{V_{\nu_{l}}}
(\theta_{I},g_{V_{\nu_{l}}},f_{1\nu_{l}}^{I},s'=s)}{d\Omega}$$
$$+\frac{d\sigma_{I}^{V_{\nu_{l}}}
(\theta_{I},g_{V_{\nu_{l}}},f_{1\nu_{l}}^{I},s'=-s)}{d\Omega}$$
$$=\rho_{I}\sigma^{I}_{o}(1-\eta^{2}_{I})^{-1}$$
\begin{equation}
\times (1+\eta_{I}^{2}tg^{2}\frac{\theta_{I}}{2})
g_{V_{\nu_{l}}}f_{1\nu_{l}}^{I}F_{I}(q_{I}^{2})
\label{22}
\end{equation}

$$\frac{d\sigma_{I}^{V_{\nu_{l}}}
(\theta_{I},g_{V_{\nu_{l}}},<r^{2}_{\nu_{l}}>_{I},s)}{d\Omega}=
\frac{d\sigma_{I}^{V_{\nu_{l}}}
(\theta_{I},g_{V_{\nu_{l}}},<r^{2}_{\nu_{l}}>_{I},s'=s)}{d\Omega}$$
$$+\frac{d\sigma_{I}^{V_{\nu_{l}}}
(\theta_{I},g_{V_{\nu_{l}}},<r^{2}_{\nu_{l}}>_{I},s'=-s)}{d\Omega}$$
$$=-\frac{2}{3}(m_{\nu_{l}}^{I})^{2}\gamma_{I}^{-1}
\rho_{I}\sigma^{I}_{o}(1-\eta^{2}_{I})^{-1}$$
\begin{equation}
\times (1+\eta_{I}^{2}tg^{2}\frac{\theta_{I}}{2})
g_{V_{\nu_{l}}}<r^{2}_{\nu_{l}}>_{I}F_{I}(q_{I}^{2})
\label{23}
\end{equation}
A purely weak interaction of partially longitudinally polarized neutrinos
is described by the cross section
$$\frac{d\sigma_{W}^{V_{\nu_{l}}}(\theta_{W},g_{V_{\nu_{l}}},s)}{d\Omega}=
\frac{d\sigma_{W}^{V_{\nu_{l}}}(\theta_{W},g_{V_{\nu_{l}}},s'=s)}{d\Omega}+
\frac{d\sigma_{W}^{V_{\nu_{l}}}(\theta_{W},g_{V_{l}},s'=-s)}{d\Omega}$$
\begin{equation}
=\frac{G_{F}^{2}(m_{\nu_{l}}^{W})^{2}}{8\pi^{2}\sin^{2}\theta_{W}}
\eta_{W}^{-2}(1+\eta_{W}^{2}tg^{2}\frac{\theta_{W}}{2})
g_{V_{\nu_{l}}}^{2}F_{W}^{2}(q_{W}^{2})\cos^{2}\frac{\theta_{W}}{2}
\label{24}
\end{equation}
Among (\ref{16})-(\ref{24}) the cross sections (\ref{18}) and (\ref{23})
have negative signs. This fall in favor of a latent connection between
the electric charge of the neutrino and its charge radius. The latter together
with (\ref{1}) permits the conclusion that at the availability of a nonzero mass,
the studied neutrino must possess simultaneously each current of a vector nature.

To define their compound structure, it is desirable to replace (\ref{10}) averaging the cross sections (\ref{11})-(\ref{13}) over $s$ and summing over $s'$ by
\begin{equation}
d\sigma_{EW}^{V_{\nu_{l}}}(\theta_{EW})=
d\sigma_{E}^{V_{\nu_{l}}}(\theta_{E})+
\frac{1}{2}d\sigma_{I}^{V_{\nu_{l}}}(\theta_{I})+
\frac{1}{2}d\sigma_{I}^{V_{\nu_{l}}}(\theta_{I})+
d\sigma_{W}^{V_{\nu_{l}}}(\theta_{W})
\label{25}
\end{equation}
Its components may be presented as
$$d\sigma_{E}^{V_{\nu_{l}}}(\theta_{E})=
d\sigma_{E}^{V_{\nu_{l}}}(\theta_{E},f_{1\nu_{l}}^{E})+
\frac{1}{2}d\sigma_{E}^{V_{\nu_{l}}}
(\theta_{E},f_{1\nu_{l}}^{E},<r^{2}_{\nu_{l}}>_{E})$$
$$+\frac{1}{2}d\sigma_{E}^{V_{\nu_{l}}}
(\theta_{E},f_{1\nu_{l}}^{E},<r^{2}_{\nu_{l}}>_{E})$$
\begin{equation}
+d\sigma_{E}^{V_{\nu_{l}}}(\theta_{E},<r^{2}_{\nu_{l}}>_{E})+
d\sigma_{E}^{V_{\nu_{l}}}(\theta_{E},f_{2\nu_{l}}^{E})
\label{26}
\end{equation}
\begin{equation}
d\sigma_{I}^{V_{\nu_{l}}}(\theta_{I})=
d\sigma_{I}^{V_{\nu_{l}}}(\theta_{I},g_{V_{\nu_{l}}},f_{1\nu_{l}}^{I})+
d\sigma_{I}^{V_{\nu_{l}}}
(\theta_{I},g_{V_{\nu_{l}}},<r^{2}_{\nu_{l}}>_{I})
\label{27}
\end{equation}
\begin{equation}
d\sigma_{W}^{V_{\nu_{l}}}(\theta_{W})=
d\sigma_{W}^{V_{\nu_{l}}}(\theta_{W},g_{V_{\nu_{l}}})
\label{28}
\end{equation}
Any part of each of (\ref{26})-(\ref{28}) coincides with the corresponding cross section 
from (\ref{16}), (\ref{21}), and (\ref{24}) and, consequently, we find
\begin{equation}
d\sigma_{E}^{V_{\nu_{l}}}(\theta_{E},f_{1\nu_{l}}^{E})=
d\sigma_{E}^{V_{\nu_{l}}}(\theta_{E},f_{1\nu_{l}}^{E},s)
\label{29}
\end{equation}
\begin{equation}
d\sigma_{E}^{V_{\nu_{l}}}
(\theta_{E},f_{1\nu_{l}}^{E},<r^{2}_{\nu_{l}}>_{E})=
d\sigma_{E}^{V_{\nu_{l}}}
(\theta_{E},f_{1\nu_{l}}^{E},<r^{2}_{\nu_{l}}>_{E},s)
\label{30}
\end{equation}
\begin{equation}
d\sigma_{E}^{V_{\nu_{l}}}(\theta_{E},<r^{2}_{\nu_{l}}>_{E})=
d\sigma_{E}^{V_{\nu_{l}}}(\theta_{E},<r^{2}_{\nu_{l}}>_{E},s)
\label{31}
\end{equation}
\begin{equation}
d\sigma_{E}^{V_{\nu_{l}}}(\theta_{E},f_{2\nu_{l}}^{E})=
d\sigma_{E}^{V_{\nu_{l}}}(\theta_{E},f_{2\nu_{l}}^{E},s)
\label{32}
\end{equation}
\begin{equation}
d\sigma_{I}^{V_{\nu_{l}}}(\theta_{I},g_{V_{\nu_{l}}},f_{1\nu_{l}}^{I})=
d\sigma_{I}^{V_{\nu_{l}}}(\theta_{I},g_{V_{\nu_{l}}},f_{1\nu_{l}}^{I},s)
\label{33}
\end{equation}
\begin{equation}
d\sigma_{I}^{V_{\nu_{l}}}
(\theta_{I},g_{V_{\nu_{l}}},<r^{2}_{\nu_{l}}>_{I})=
d\sigma_{I}^{V_{\nu_{l}}}
(\theta_{I},g_{V_{\nu_{l}}},<r^{2}_{l}>_{\nu_{l}},s)
\label{34}
\end{equation}
\begin{equation}
d\sigma_{W}^{V_{\nu_{l}}}(\theta_{W},g_{V_{\nu_{l}}})=
d\sigma_{W}^{V_{\nu_{l}}}(\theta_{W},g_{V_{\nu_{l}}},s)
\label{35}
\end{equation}
It is already clear from them that (\ref{10}) describes the scattering of
a partially ordered flux of unpolarized and longitudinal polarized fermions.
It can therefore constitute [6,7] a kind of set of cross sections
\begin{equation}
d\sigma_{EW}^{V_{\nu_{l}}}=
\{d\sigma_{EW}^{V_{\nu_{l}}}(\theta_{EW},s), \, \, \, \,
d\sigma_{EW}^{V_{\nu_{l}}}(\theta_{EW})\}.
\label{36}
\end{equation}
The compound structure of both elements of (\ref{36}) testifies that
any of (\ref{15}) and (\ref{25}) constitute the naturally united subclass:
$$d\sigma_{EW}^{V_{\nu_{l}}}(\theta_{EW},s)=
\{d\sigma_{E}^{V_{\nu_{l}}}(\theta_{E},f_{1\nu_{l}}^{E},s),  \, \, \, \,
\frac{1}{2}d\sigma_{E}^{V_{\nu_{l}}}
(\theta_{E},f_{1\nu_{l}}^{E},<r^{2}_{\nu_{l}}>_{E},s),$$
$$\frac{1}{2}d\sigma_{E}^{V_{\nu_{l}}}
(\theta_{E},f_{1\nu_{l}}^{E},<r^{2}_{\nu_{l}}>_{E},s), 
\, \, \, \,
d\sigma_{E}^{V_{\nu_{l}}}(\theta_{E},<r^{2}_{\nu_{l}}>_{E},s),$$
$$d\sigma_{E}^{V_{\nu_{l}}}(\theta_{E},f_{2\nu_{l}}^{E},s),
\, \, \, \,
\frac{1}{2}d\sigma_{I}^{V_{\nu_{l}}}
(\theta_{I},g_{V_{\nu_{l}}},f_{1\nu_{l}}^{I},s),$$
$$\frac{1}{2}d\sigma_{I}^{V_{\nu_{l}}}
(\theta_{I},g_{V_{\nu_{l}}},f_{1\nu_{l}}^{I},s),
\, \, \, \,
\frac{1}{2}d\sigma_{I}^{V_{\nu_{l}}}
(\theta_{I},g_{V_{\nu_{l}}},<r^{2}_{\nu_{l}}>_{I},s),$$
\begin{equation}
\frac{1}{2}d\sigma_{I}^{V_{\nu_{l}}}
(\theta_{I},g_{V_{\nu_{l}}},<r^{2}_{\nu_{l}}>_{I},s),
\, \, \, \,
d\sigma_{W}^{V_{\nu_{l}}}(\theta_{W},g_{V_{\nu_{l}}},s)\}
\label{37}
\end{equation}
$$d\sigma_{EW}^{V_{\nu_{l}}}(\theta_{EW})=
\{d\sigma_{E}^{V_{\nu_{l}}}(\theta_{E},f_{1\nu_{l}}^{E}),
\, \, \, \,
\frac{1}{2}d\sigma_{E}^{V_{\nu_{l}}}
(\theta_{E},f_{1\nu_{l}}^{E},<r^{2}_{\nu_{l}}>_{E}),$$
$$\frac{1}{2}d\sigma_{E}^{V_{\nu_{l}}}
(\theta_{E},f_{1\nu_{l}}^{E},<r^{2}_{\nu_{l}}>_{E}),
\, \, \, \,
d\sigma_{E}^{V_{\nu_{l}}}(\theta_{E},<r^{2}_{\nu_{l}}>_{E}),$$
$$d\sigma_{E}^{V_{\nu_{l}}}(\theta_{E},f_{2\nu_{l}}^{E}), \, \, \, \,
\frac{1}{2}d\sigma_{I}^{V_{\nu_{l}}}
(\theta_{I},g_{V_{\nu_{l}}},f_{1\nu_{l}}^{I}),$$
$$\frac{1}{2}d\sigma_{I}^{V_{\nu_{l}}}
(\theta_{I},g_{V_{\nu_{l}}},f_{1\nu_{l}}^{I}),
\, \, \, \,
\frac{1}{2}d\sigma_{I}^{V_{\nu_{l}}}
(\theta_{I},g_{V_{\nu_{l}}},<r^{2}_{\nu_{l}}>_{I}),$$
\begin{equation}
\frac{1}{2}d\sigma_{I}^{V_{\nu_{l}}}
(\theta_{I},g_{V_{\nu_{l}}},<r^{2}_{\nu_{l}}>_{I}),
\, \, \, \,
d\sigma_{W}^{V_{\nu_{l}}}(\theta_{W},g_{V_{\nu_{l}}})\}
\label{38}
\end{equation}
These subsets, according to (\ref{29})-(\ref{35}), must have the same size. This implies that 
their elements correspond in nature to one of the previously mentioned difermions, because of 
which, all components of cross sections (\ref{15}) and (\ref{25}) coincide.

Another important circumstance is that between the fermions of
each of paraparticles (\ref{14}) there exists a sharp flavor symmetrical
dependence [20]. Such a connection gives the right to use the flavor
symmetry as a theorem [6,7] about the equality of the structural parts of
cross sections of the neutrino interaction with vector electroweak currents.

\vspace{0.8cm}
\noindent
{\bf 3. Mass structure of neutrinos of a vector nature}
\vspace{0.4cm}

The preceding reasoning says that the possible pairs of elements from (\ref{36}) establish 42 ratios. Jointly with the expressions of cross sections (\ref{29})-(\ref{35}), the latter are reduced to 21 explicit equations.

To show their structural picture, it is sufficient to choose five from the starting relations
\begin{equation}
\frac{d\sigma_{W}^{V_{\nu_{l}}}(\theta_{W},g_{V_{\nu_{l}}})}
{d\sigma_{E}^{V_{\nu_{l}}}(\theta_{E},f_{i\nu_{l}}^{E})}=1
\, \, \, \,
\frac{2d\sigma_{W}^{V_{\nu_{l}}}(\theta_{W},g_{V_{\nu_{l}}})}
{d\sigma_{I}^{V_{\nu_{l}}}(\theta_{I},g_{V_{\nu_{l}}},f_{1\nu_{l}}^{I})}=1
\label{39}
\end{equation}
\begin{equation}
\frac{d\sigma_{W}^{V_{\nu_{l}}}(\theta_{W},g_{V_{\nu_{l}}})}
{d\sigma_{E}^{V_{\nu_{l}}}(\theta_{E},<r^{2}_{\nu_{l}}>_{E})}=1
\, \, \, \,
\frac{2d\sigma_{W}^{V_{\nu_{l}}}(\theta_{W},g_{V_{\nu_{l}}})}
{d\sigma_{I}^{V_{\nu_{l}}}
(\theta_{I},g_{V_{\nu_{l}}},<r^{2}_{\nu_{l}}>_{I})}=1
\label{40}
\end{equation}
It is not excluded, however, that the discussed processes depend [25] not only
on the fermion properties but also on the structure of a nucleus itself.

For elucidation of the nature of the neutrino, it is desirable to use a nucleus
with zero spin and isospin. Therefore, if $N=Z$ then inserting the exact values
of cross sections from (\ref{29})-(\ref{35}) into (\ref{39}) and (\ref{40}), it
is not difficult to constitute those equalities that at large energies
$(E_{\nu_{l}}^{K}\gg m_{\nu_{l}}^{K})$ when
$$lim_{\eta_{K}\rightarrow 0,\theta_{K}\rightarrow 0}
\frac{\eta^{2}_{K}}{(1-\eta^{2}_{K})
\sin^{2}(\theta_{K}/2)}=-2$$
$$lim_{\eta_{E}\rightarrow 0,\theta_{E}\rightarrow 0}
\frac{\eta^{2}_{E}\sin^{-2}(\theta_{E}/2)}
{(1+\eta_{E}^{2}tg^{2}(\theta_{E}/2))
\cos^{2}(\theta_{E}/2)}=4$$
lead us to a system
\begin{equation}
f_{1\nu_{l}}^{E}(0)=-g_{V_{\nu_{l}}}
\frac{G_{F}m_{\nu_{l}}^{E}m_{\nu_{l}}^{W}}{\pi\sqrt{2}\alpha}\sin\theta_{W}
\label{41}
\end{equation}
\begin{equation}
f_{2\nu_{l}}^{E}(0)=-g_{V_{\nu_{l}}}
\frac{G_{F}m_{\nu_{l}}^{W}}{2\pi\sqrt{2}\alpha}\sin\theta_{W}
\label{42}
\end{equation}
\begin{equation}
<r^{2}_{\nu_{l}}>_{E}=-g_{V_{\nu_{l}}}
\frac{3G_{F}}{\pi\sqrt{2}\alpha}
\frac{m_{\nu_{l}}^{W}}{m_{\nu_{l}}^{E}}\sin\theta_{W}
\label{43}
\end{equation}
\begin{equation}
f_{1\nu_{l}}^{I}(0)=-g_{V_{\nu_{l}}}
\frac{G_{F}(m_{\nu_{l}}^{W})^{2}}{\pi\sqrt{2}\alpha}\sin\theta_{W}
\label{44}
\end{equation}
\begin{equation}
<r^{2}_{\nu_{l}}>_{I}=-g_{V_{\nu_{l}}}\frac{3G_{F}}{\pi\sqrt{2}\alpha}
\left(\frac{m_{\nu_{l}}^{W}}{m_{\nu_{l}}^{I}}\right)^{2}\sin\theta_{W}
\label{45}
\end{equation}

Comparing (\ref{41})-(\ref{45}), it is easy to observe the characteristic
dependence of each of $f_{1\nu_{l}}(0)$ and $<r^{2}_{\nu_{l}}>$ on the sizes
of $m_{\nu_{l}}^{E}$ and $m_{\nu_{l}}^{W},$ which may serve as an indication
to the existence of both types of masses.

Thus, we have the possibility on the basis of (\ref{41}) and its logical prediction about 
\begin{equation}
e_{\nu_{l}}^{E}=-g_{V_{\nu_{l}}}
\frac{G_{F}m_{\nu_{l}}^{E}m_{\nu_{l}}^{W}}{\pi\sqrt{2}\alpha}\sin\theta_{W}
\label{46}
\end{equation}
to establish the mass picture of neutrinos of vector currents
in a latent united form
\begin{equation}
f_{1\nu_{l}}^{E}(0)=e_{\nu_{l}}^{E}
\label{47}
\end{equation}
\begin{equation}
f_{2\nu_{l}}^{E}(0)=\frac{e_{\nu_{l}}^{E}} {2m_{\nu_{l}}^{E}}
\label{48}
\end{equation}
\begin{equation}
<r^{2}_{\nu_{l}}>_{E}=\frac{3e_{\nu_{l}}^{E}}{ (m_{\nu_{l}}^{E})^{2}}
\label{49}
\end{equation}
\begin{equation}
f_{1\nu_{l}}^{I}(0)=\frac{m_{\nu_{l}}^{W}}{m_{\nu_{l}}^{E}}e_{\nu_{l}}^{E}
\label{50}
\end{equation}
\begin{equation}
<r^{2}_{\nu_{l}}>_{I}=\frac{m_{\nu_{l}}^{E}m_{\nu_{l}}^{W}}
{(m_{\nu_{l}}^{I})^{2}}<r^{2}_{\nu_{l}}>_{E}
\label{51}
\end{equation}
Unification of (\ref{47}) and (\ref{48}) suggests a connection
\begin{equation}
f_{1\nu_{l}}^{E}(0)-2m_{\nu_{l}}^{E}f_{2\nu_{l}}^{E}(0)=0
\label{52}
\end{equation}

Its comparison with (\ref{1}) convinces us here that any of all types of charges 
leads to the appearance of a kind of dipole moment [14].

\vspace{0.8cm}
\noindent
{\bf 4. Anomalous behavior of neutrinos of vector currents}
\vspace{0.4cm}

It is already clear from the preceding reasoning that (\ref{41})-(\ref{43}) give the 
normal charge, charge radius, and magnetic moment: 
$e_{\nu_{l}}^{norm}=f_{1\nu_{l}}^{E}(0),$
$\mu_{\nu_{l}}^{norm}=f_{2\nu_{l}}^{E}(0),$
$<r^{2}_{\nu_{l}}>_{norm}=<r^{2}_{\nu_{l}}>_{E}.$

Furthermore, if we suppose that in the case of a neutrino, the Schwinger
value of magnetic moment [26] has an estimate $\mu_{\nu_{l}}^{anom}=
(\alpha/2\pi)\mu_{\nu_{l}}^{norm},$ in a similar way one can get from
(\ref{41})-(\ref{43}) the following functions:
\begin{equation}
e_{\nu_{l}}^{anom}=-g_{V_{\nu_{l}}}
\frac{G_{F}m_{\nu_{l}}^{E}m_{\nu_{l}}^{W}}{2\pi^{2}\sqrt{2}}\sin\theta_{W}
\label{53}
\end{equation}
\begin{equation}
\mu_{\nu_{l}}^{anom}=-g_{V_{\nu_{l}}}
\frac{G_{F}m_{\nu_{l}}^{W}}{4\pi^{2}\sqrt{2}}\sin\theta_{W}
\label{54}
\end{equation}
\begin{equation}
<r^{2}_{\nu_{l}}>_{anom}=-g_{V_{\nu_{l}}}\frac{3G_{F}}{2\pi^{2}\sqrt{2}}
\frac{m_{\nu_{l}}^{W}}{m_{\nu_{l}}^{E}}\sin\theta_{W}
\label{55}
\end{equation}
The absence of one of the components of mass would imply that the mass itself does not exist at 
all. Nevertheless, if we consider the case when $m_{\nu_{l}}^{E}=m_{\nu_{l}}^{W}=m_{\nu_{l}},$ (\ref{53})-(\ref{55}) take the form
\begin{equation}
e_{\nu_{l}}^{anom}=-g_{V_{\nu_{l}}}
\frac{G_{F}m_{\nu_{l}}^{2}}{2\pi^{2}\sqrt{2}}\sin\theta_{W}
\label{56}
\end{equation}
\begin{equation}
\mu_{\nu_{l}}^{anom}=-g_{V_{\nu_{l}}}
\frac{G_{F}m_{\nu_{l}}}{4\pi^{2}\sqrt{2}}\sin\theta_{W}
\label{57}
\end{equation}
\begin{equation}
<r^{2}_{\nu_{l}}>_{anom}=-g_{V_{\nu_{l}}}
\frac{3G_{F}}{2\pi^{2}\sqrt{2}}\sin\theta_{W}
\label{58}
\end{equation}
It has been mentioned earlier that our implications refer to any Dirac neutrino of a
C-invariant nature. This gives the right to apply to the case when $g_{V_{\nu_{l}}}=1,$ 
and $\beta_{V}^{(0)}=1.$ At such a choice of constants, (\ref{39}) and (\ref{40}) replace (\ref{53})-(\ref{55}) by
\begin{equation}
e_{\nu_{l}}^{anom}=\frac{G_{F}m_{\nu_{l}}^{2}}{4\pi^{2}\sqrt{2}}
\label{59}
\end{equation}
\begin{equation}
\mu_{\nu_{l}}^{anom}=\frac{G_{F}m_{\nu_{l}}}{8\pi^{2}\sqrt{2}}
\label{60}
\end{equation}
\begin{equation}
<r^{2}_{\nu_{l}}>_{anom}=\frac{3G_{F}}{4\pi^{2}\sqrt{2}}
\label{61}
\end{equation}
The latter together with $e_{\nu_{l}}^{norm},$ $\mu_{\nu_{l}}^{norm},$ and $<r^{2}_{\nu_{l}}>_{norm}$ permit finding the full charge, charge radius, and magnetic moment:
$e_{\nu_{l}}^{full}=(1+\alpha/2\pi)e_{\nu_{l}}^{norm},$
$\mu_{\nu_{l}}^{full}=(1+\alpha/2\pi)\mu_{\nu_{l}}^{norm},$
$<r^{2}_{\nu_{l}}>_{full}=(1+\alpha/2\pi)<r^{2}_{\nu_{l}}>_{norm}.$

The basis for such an approach is that from the point of view of a Dirac particle itself,
$\mu_{\nu_{l}}^{anom}$ can exist in the presence of the anomalous charge [14] having 
a kind of radius.

\vspace{0.8cm}
\noindent
{\bf 5. Neutrino universality}
\vspace{0.4cm}

We recognize that (\ref{41})-(\ref{52}) remain valid also for all types of leptons, and 
the constants $g_{V_{\nu_{l}}}$ and $g_{V_{l}}$ have the same value. Then it is possible, for example, to relate (\ref{41})-(\ref{43}) to a renormalized size [27]
\begin{equation}
e_{e}^{E}=-g_{V_{e}}
\frac{G_{F}m_{e}^{E}m_{e}^{W}}{\pi\sqrt{2}\alpha}\sin\theta_{W}
\label{62}
\end{equation}
owing to which, they are expressed in units of the electron charge $e_{e}^{E}$ and Bohr 
magnetons $\mu_{B}=e_{e}^{E}/2m_{e}^{E}$ in the following manner:
\begin{equation}
e_{\nu_{l}}^{E}=f_{1\nu_{l}}^{E}(0)=\frac{m_{\nu_{l}}^{E}}{m_{e}^{E}}
\frac{m_{\nu_{l}}^{W} }{m_{e}^{W}}e_{e}^{E}
\label{63}
\end{equation}
\begin{equation}
\mu_{\nu_{l}}^{E}=
f_{2\nu_{l}}^{E}(0)=\frac{m_{\nu_{l}}^{W}}{m_{e}^{W}}\mu_{B}
\label{64}
\end{equation}
\begin{equation}
<r^{2}_{\nu_{l}}>_{E}=
\frac{m_{\nu_{l}}^{W}}{m_{e}^{W}}
\frac{3e_{e}^{E}}{m_{\nu_{l}}^{E}m_{e}^{E}}
\label{65}
\end{equation}
Turning again to (\ref{41}) and (\ref{42}), we remark that their interratio
for any lepton and its neutrino coincide. This may serve as an indication
of the existence of a relation between the fermion masses
\begin{equation}
\frac{m_{\nu_{l}}^{E}}{m_{l}^{E}}=\frac{f_{1\nu_{l}}^{E}(0)}{f_{1l}^{E}(0)}
\frac{f_{2l}^{E}(0)}{f_{2\nu_{l}}^{E}(0)}
\label{66}
\end{equation}
It together with the full lepton number conservation predicts the size
of the neutrino electric mass
\begin{equation}
m_{\nu_{e}}^{E}:m_{\nu_{\mu}}^{E}:m_{\nu_{\tau}}^{E}=
m_{e}^{E}:m_{\mu}^{E}:m_{\tau}^{E}
\label{67}
\end{equation}
In the same way one can find from (\ref{42}) and (\ref{44}) that
\begin{equation}
m_{\nu_{e}}^{W}:m_{\nu_{\mu}}^{W}:m_{\nu_{\tau}}^{W}=
m_{e}^{W}:m_{\mu}^{W}:m_{\tau}^{W}
\label{68}
\end{equation}
So it is seen that the electric and weak masses of the neutrino are
proportional to the electric and weak masses, respectively, of a lepton of
the same family of doublets. They establish one more highly important ratio
\begin{equation}
m_{\nu_{e}}^{E}m_{\nu_{e}}^{W}:m_{\nu_{\mu}}^{E}m_{\nu_{\mu}}^{W}:
m_{\nu_{\tau}}^{E}m_{\nu_{\tau}}^{W}=
m_{e}^{E}m_{e}^{W}:m_{\mu}^{E}m_{\mu}^{W}:m_{\tau}^{E}m_{\tau}^{W}
\label{69}
\end{equation}
Therefore, if it turns out that lepton universality expresses [27] the idea of
a constant size law  
\begin{equation}
m_{l}^{E}m_{l}^{W}=const
\label{70}
\end{equation}
from the point of view of each mass formula (\ref{69}) or (\ref{70}),
it should be expected that the availability of a connection
\begin{equation}
m_{\nu_{l}}^{E}m_{\nu_{l}}^{W}=const
\label{71}
\end{equation}
requires one to follow the logic of neutrino universality in the mass structure
dependence of neutrinos.

\vspace{0.8cm}
\noindent
{\bf 6. Conclusion}
\vspace{0.4cm}

Analysis of experimental results [28-30] assumed that
$$\mu_{\nu_{e}}^{E}< 0.74\cdot 10^{-10}\ {\rm \mu_{B}} \, \, \, \,
\mu_{\nu_{\mu}}^{E}< 6.8\cdot 10^{-10}\ {\rm \mu_{B}} \, \, \, \,
\mu_{\nu_{\tau}}^{E}< 3.9\cdot 10^{-7}\ {\rm \mu_{B}}$$
Having (\ref{64}) and taking into account [27] that
$m_{e}^{W}=5.15\cdot 10^{-2}\ {\rm eV},$
we establish here the first estimates of the neutrino weak masses
$$m_{\nu_{e}}^{W}< 3.81\cdot 10^{-12}\ {\rm eV}$$
$$m_{\nu_{\mu}}^{W}< 3.5\cdot 10^{-11}\ {\rm eV}$$
$$m_{\nu_{\tau}}^{W}< 2.08\cdot 10^{-8}\ {\rm eV}$$
Known laboratory data [31-33] for the neutrino rest mass lead
to the following restrictions:
$$m_{\nu_{e}}^{E}< 2.5\ {\rm eV} \, \, \, \,
m_{\nu_{\mu}}^{E}< 0.17\ {\rm MeV} \, \, \, \,
m_{\nu_{\tau}}^{E}< 18.2\ {\rm MeV}$$
Insertion of $m_{e}^{E}$ and $m_{e}^{W}$ into (\ref{63}) at these values
of $m_{\nu_{l}}^{E}$ and $m_{\nu_{l}}^{W}$ gives
$$e_{\nu_{e}}^{E}< 3.62\cdot 10^{-16}\ {\rm e_{e}^{E}}$$
$$e_{\nu_{\mu}}^{E}< 2.26\cdot 10^{-10}\ {\rm e_{e}^{E}}$$
$$e_{\nu_{\tau}}^{E}< 1.44\cdot 10^{-5}\ {\rm e_{e}^{E}}$$
The size of $e_{\nu_{\mu}}^{E}$ may be accepted as a new estimate.
The values of $e_{\nu_{e}}^{E}$ and $e_{\nu_{\tau}}^{E}$ are compatible
with those that follow from experiments [34,35] 
$$e_{\nu_{e}}^{E}< 2\cdot 10^{-15}\ {\rm e_{e}^{E}} \, \, \, \,
e_{\nu_{\tau}}^{E}< 4\cdot 10^{-4}\ {\rm e_{e}^{E}}$$
One can also find from (\ref{65}) that
$$<r^{2}_{\nu_{e}}>_{E}< 2.78\cdot 10^{-35}\ {\rm cm^{2}}$$
$$<r^{2}_{\nu_{\mu}}>_{E}< 3.76\cdot 10^{-39}\ {\rm cm^{2}}$$
$$<r^{2}_{\nu_{\tau}}>_{E}< 2.09\cdot 10^{-38}\ {\rm cm^{2}}$$
where $<r^{2}_{\nu_{\tau}}>_{E}$ must be interpreted as the
measured charge radius for the first time, $<r^{2}_{\nu_{e}}>_{E}$ and
$<r^{2}_{\nu_{\mu}}>_{E}$ essentially improve facts already available
in the literature [30,36]
$$<r^{2}_{\nu_{e}}>_{E}< 4.14\cdot 10^{-32}\ {\rm cm^{2}} \, \, \, \,
<r^{2}_{\nu_{\mu}}>_{E}< 0.68\cdot 10^{-32}\ {\rm cm^{2}}$$
At first sight, because of the difference dependence in masses, all neutrinos have
neither an equal charge, the same charge radius, nor an identical magnetic 
moment. There exist, however, earlier [37] and comparatively 
recent [38] laboratory restrictions on the size of $e_{\nu_{l}}^{E},$ 
which may testify in favor of neutrino universality. It follows from  
ref. 37 in addition that 
$e_{\nu_{l}}^{E}< 2\cdot 10^{-13}\ {\rm e_{e}^{E}}.$
This value together with (\ref{63}) or (\ref{71}) gives the possibility to directly 
look on the nature of weak masses of the vector types of Dirac neutrinos
$$m_{\nu_{e}}^{W}< 2.1\cdot 10^{-9}\ {\rm eV}$$
$$m_{\nu_{\mu}}^{W}< 3.09\cdot 10^{-14}\ {\rm eV}$$
$$m_{\nu_{\tau}}^{W}< 2.89\cdot 10^{-16}\ {\rm eV}$$
Thus, the existence both of an electric and a weak component of mass is by no means excluded experimentally.

In the mass type dependence, a neutrino has a nonzero charge, charge radius,
and magnetic moment. Therefore, it is not surprising that the previously mentioned
experiments may serve as a practical confirmation of the availability of
universal mass structure of the interaction with any of the corresponding
types of gauge bosons of all Dirac neutrinos of vector currents. Of course,
an observed regularity reflects the sharply expressed features of mass and charge
and thereby opens the chance for creation of the unified picture of nature
of elementary particles and fields.

\vspace{0.8cm}
\noindent
{\bf References}
\begin{enumerate}
\item
B. Pontecorvo. J. Exp. Theor. Phys. {\bf 33}, 549 (1957).
\item
E. Majorana. Il Nuovo Cimento, {\bf 14}, 171 (1937). doi:10.1007/BF02961314.
\item
R.S. Sharafiddinov. Phys. Essays, {\bf 19}, 58 (2006). hep-ph/0407262. 
doi:10.4006/1.3025781.
\item
Y. Fukuda, T. Hayakawa, E. Ichihara, et al. Phys. Rev. Lett. {\bf 81}, 1562 (1998).

doi:10.1103/PhysRevLett.81.1562.
\item
W. Wang. AIP Conf. Proc. {\bf 1222}, 494 (2010). 0910.4605 [hep-ex].
\item
R.S. Sharafiddinov. In Proceedings of the April Meeting of the American
Physical Society, Jacksonville, Florida, 14–17 April 2007. Abstract, K11.00008.
\item
R.S. Sharafiddinov. J. Phys. Nat. Sci. {\bf 4}, 1 (2013). physics/0702233.
\item
R.S. Sharafiddinov. In Proceedings of the April Meeting of the American
Physical Society, Dallax, Texas, 22–25 April 2006. Abstract, D1.00076.
\item
R.S. Sharafiddinov. Fizika B, {\bf 16}, 1 (2007). hep-ph/0512346.
\item
E. Fermi Rend. Lincei, {\bf 31}, 184, 306 (1922).
\item
E. Fermi. Phys. Zeit. {\bf 23}, 340 (1922).
\item
T.D. Lee and C.N. Yang. Phys. Rev. {\bf 105}, 1671 (1957). 

doi:10.1103/PhysRev.105.
\item
T.D. Lee and C.S. Wu. Annu. Rev. Nucl. Sci. {\bf 15}, 381 (1965). 

doi:10.1146/annurev.ns.15.120165.002121.
\item
R.S. Sharafiddinov. Spacetime Subst. {\bf 3}, 86 (2002). physics/0305009.
\item
R.S. Sharafiddinov. Spacetime Subst. {\bf 3}, 132 (2002). physics/0305014.
\item
I.S. Batkin and M.K. Sundaresan. J. Phys. G, {\bf 20}, 1749 (1994). 

doi:10.1088/0954-3899/20/11/004.
\item
S.L. Glashow. Nucl. Phys. {\bf 22}, 579 (1961). doi:10.1016/0029-5582(61)90469-2.
\item
A. Salam and J.C. Ward. Phys. Lett. {\bf 13}, 168 (1964). 
doi:10.1016/0031-9163(64)90711-5.
\item
S. Weinberg. Phys. Rev. Lett. {\bf 19}, 1264 (1967). doi:10.1103/PhysRevLett.19.1264.
\item
R.S. Sharafiddinov. In Proceedings of the April Meeting of the American
Physical Society, Dallax, Texas, 22–25 April 2006. Abstract, H12.00009. 
arXiv:hep-ph/0511065.
\item
R.S. Sharafiddinov. Spacetime Subst. {\bf 3}, 47 (2002). physics/0305008.
\item
S. Weinberg. Phys. Rev. Lett. {\bf 29}, 388 (1972). doi:10.1103/PhysRevLett.29.388.
\item
R.S. Sharafiddinov. Spacetime Subst. {\bf 4}, 87 (2003). hep-ph/0401230.
\item
R.S. Sharafiddinov. Spacetime Subst. {\bf 3}, 134 (2002). physics/0305015.
\item
R.B. Begzhanov and R.S. Sharafiddinov. In Proceedings of the International
Conference on Nuclear Physics, Moscow, 16–19 June 1998. St-Petersburg.
Abstracts, p. 354.
\item
J. Schwinger. Phys. Rev. {\bf 76}, 790 (1949). doi:10.1103/PhysRev.76.790.
\item
R.S. Sharafiddinov. Eur. Phys. J. Plus, {\bf 126}, 40 (2011). 
arXiv:0802.3736 [physics.gen-ph]. doi:10.1140/epjp/i2011-11040-x.
\item
H.T. Wong, H. Li, S. Lin, et al. Phys. Rev. D, {\bf 75}, 012001 (2007). 

doi:10.1103/PhysRevD.75.012001.
\item
R. Schwienhorst, D. Ciampa, C. Erickson, et al. Phys. Lett. B, {\bf 513}, 23 (2001).
doi:10.1016/S0370-2693(01)00746-8.
\item
L.B. Auerbach, R. Burman, D. Caldwell, et al. Phys. Rev. D, {\bf 63}, 112001 (2001).

doi:10.1103/PhysRevD.63.112001.
\item
V.M. Lobashev, V.N. Aseev, A.I. Belesev, et al. Phys. Lett. B, {\bf 460}, 227 (1999).
doi:10.1016/S0370-2693(99)00781-9.
\item
K.A. Assamagan, et al. Phys. Rev. D, {\bf 53}, 6065 (1996). 

doi:10.1103/PhysRevD.53.6065.
\item
R. Barate, et al. Eur. Phys. J. C, {\bf 2}, 395 (1998). 

doi:10.1007/s100520050149.
\item
G. Barbiellini and G. Cocconi. Nature, {\bf 329}, 21 (1987). 

doi:10.1038/329021b0.PMID:3041224.
\item
K.S. Babu, T.M. Gould, and I.Z. Rothstein. Phys. Lett. B, {\bf 321}, 140 (1994). 

doi:10.1016/0370-2693(94)90340-9.
\item
M. Hirsch, E. Nardi, and D. Restrepo. Phys. Rev. D, {\bf 67}, 033005 (2003). 

doi:10.1103/PhysRevD.67.033005.
\item
J. Bernstein, M. Ruderman, and G. Feinberg. Phys. Rev. {\bf 132}, 1227 (1963).

doi:10.1103/PhysRev.132.1227.
\item
G.G. Raffelt. Phys. Rep. {\bf 320}, 319 (1999). 
doi:10.1016/S0370-1573(99)00074-5.
\end{enumerate}
\end{document}